# Status epilepticus and thinning of the entorhinal cortex

Jonathan Horsley[1], Yujiang Wang[1], Callum Simpson[1], Vyte Janiukstyte[1], Karoline Leiberg[1], Beth Little[1], Jane de Tisi[3], John Duncan[3], Peter N. Taylor[1,2,3*]

1. CNNP Lab (www.cnnp-lab.com), Interdisciplinary Computing and Complex BioSystems Group, School of Computing, Newcastle University, Newcastle upon Tyne, United Kingdom

2. Faculty of Medical Sciences, Newcastle University, Newcastle upon Tyne, United Kingdom

3. UCL Queen Square Institute of Neurology, Queen Square, London, United Kingdom

* Peter.Taylor@newcastle.ac.uk

## Abstract

Status epilepticus (SE) carries risks of morbidity and mortality. Experimental studies have implicated the entorhinal cortex in prolonged seizures; however, studies in large human cohorts are limited. We hypothesised that individuals with temporal lobe epilepsy (TLE) and a history of SE would have more severe entorhinal atrophy compared to others with TLE and no history of SE .

357 individuals with drug resistant temporal lobe epilepsy (TLE) and 100 healthy controls were scanned on a 3T MRI. For all subjects the cortex was segmented, parcellated, and the thickness calculated from the T1-weighted anatomical scan. Subcortical volumes were derived similarly. Cohen's d and Wilcoxon rank-sum tests respectively were used to capture effect sizes and significance.

Individuals with TLE and SE had reduced entorhinal thickness compared to those with TLE and no history of SE. The entorhinal cortex was more atrophic ipsilaterally (d=0.51, p<0.001) than contralaterally (d=0.37, p=0.01). Reductions in ipsilateral entorhinal thickness were present in both left TLE (n=22:176, d=0.78, p<0.001), and right TLE (n=19:140, d=0.31, p=0.04), albeit with a smaller effect size in right TLE. Several other regions exhibited atrophy in individuals with TLE, but these did not relate to a history of SE.

These findings suggest potential involvement or susceptibility of the entorhinal cortex in prolonged seizures.

# Introduction

Status epilepticus (SE) is relatively common in individuals with drug resistant epilepsy[1]. SE carries risks of premature mortality, morbidity, and may result in atrophy[2,3].

In temporal lobe epilepsy (TLE), the most common form of drug resistant focal epilepsy, abnormalities visible on MRI have been noted, including in the entorhinal cortex[4]. The entorhinal cortex (Figure 1) has strong reciprocal connections to mesial temporal structures including the hippocampus and piriform cortex[5]. Experimental studies in rats have demonstrated neuronal degeneration in the entorhinal cortex after SE[6], and postulated enhanced excitability in the entorhinal cortex following SE[7]. These studies suggest involvement of entorhinal cortex in, or following SE, which may be evident in neuroimaging.

Here we investigate structural abnormalities, specifically cortical thickness and subcortical volume, in a large cohort of 357 individuals with drug resistant TLE, 11% of whom had a history of SE. We hypothesised that brain regions integral to SE would have more substantial brain abnormalities in those individuals with a history of SE.

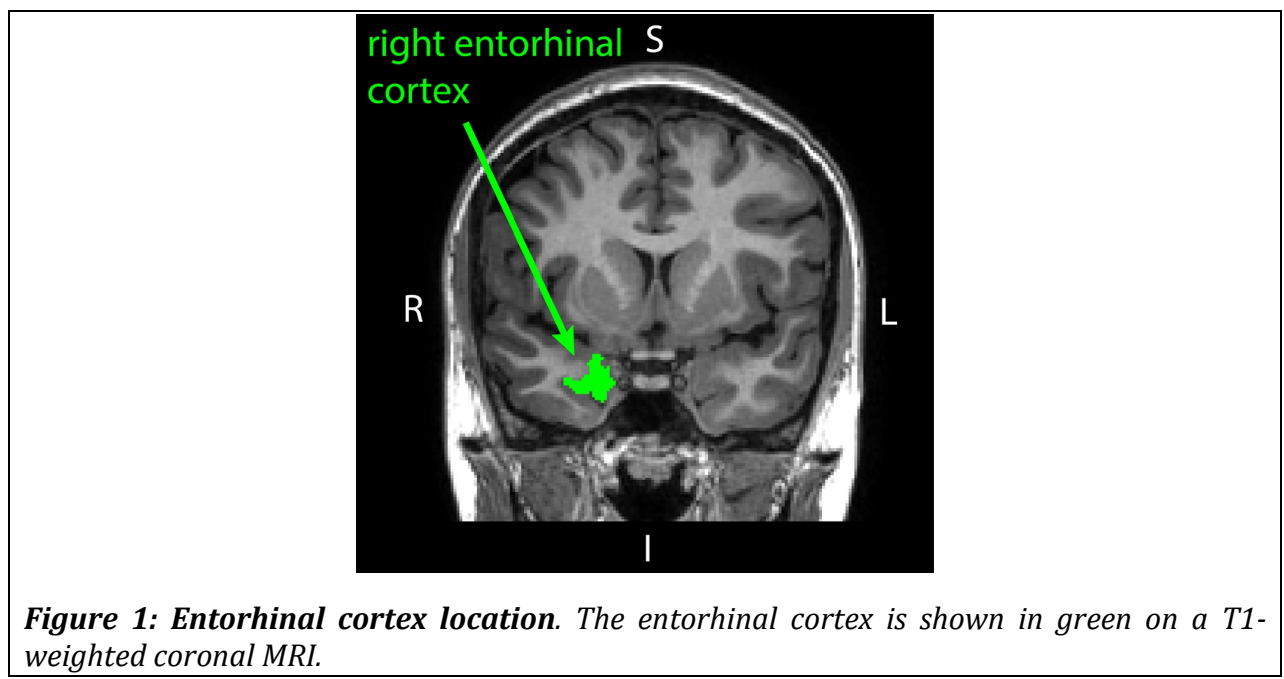

*Figure 1: Entorhinal cortex location*. *The entorhinal cortex is shown in green on a T1-weighted coronal MRI.*

# Methods

## Patient information

We identified 357 individuals with TLE who subsequently had temporal lobe surgery at the National Hospital for Neurology & Neurosurgery, London, UK between January 2003 and June 2022. All individuals underwent a pre-operative T1-weighted MRI scan on a 3T GE scanner. The cohort analysed here are a subset of those in the IDEAS dataset, in which detailed acquisition parameters have been previously shared[8]. Of the 357 patients, 41 (11%) had a history of SE. In this cohort, an individual was defined as having SE if they experienced at least one occurrence of generalised convulsive SE with a duration of > 30 minutes. There were no differences in age, sex or lateralisation of epilepsy between those with and without a history of SE (Table 1). This study of anonymized data that had been previously acquired was approved by the Health Research Authority, without the necessity to obtain individual subject consent (UCLH epilepsy surgery database: 22/SC/0016).

*Table 1: Patient demographic data, stratified by SE. Sex and side differences were assessed using Chi-squared tests. Age differences were assessed using a Wilcoxon rank-sum test.*

|                    | SE           | not SE       | Test statistic              |
|-------------------:|:------------:|:------------:|:---------------------------:|
| n                  | 41           | 316          |                             |
| Age, median (IQR)  | 35.1 (14.1)  | 35.3 (16.3)  | $W = 6773, p = 0.64$        |
| Sex, male:female   | 18:23        | 140:176      | $\chi^2 \approx 0.000, p = 1$ |
| Side, left:right   | 22:19        | 176:140      | $\chi^2 = 0.006, p = 0.94$  |

## Statistical analysis

Full details of image processing and statistical analysis have been described previously[8]. Briefly, pre-operative T1 weighted scans were run through the FreeSurfer segmentation and parcellation pipeline 'recon-all' and quality checked[4,9]. For each region in the Desikan-Killiany atlas, subcortical volumes and neocortical thicknesses were then extracted. Different scanning parameters can impact these regional measures[10], so ComBat was used to remove the scanner differences, whilst preserving biological variability[11].

We accounted for the effects of sex and healthy ageing in each region using normative modelling. To do this, we derived expected values of cortical thickness and subcortical volume given a subject's age and sex using a generalised additive model (GAM). The residual (i.e. difference) between the actual and expected values was used to calculate regional abnormalities by z-scoring against the 100 healthy controls. These abnormalities quantify how many standard deviations a subject's regional thickness or volume is away from a control population, accounting for the aforementioned age and sex effects.

We quantified the extent of alterations in cortical thickness or subcortical volume across the cohort using Cohen's d. In each region, we calculated the effect size of the difference in abnormality between those individuals with SE, compared to those without SE. We tested these differences for significance using the Wilcoxon rank-sum test.

## Results

### Status epilepticus is associated with thinning of the entorhinal cortex

Compared to those without a history of SE, individuals with SE had reduced cortical thickness in the ipsilateral entorhinal cortex (Figure 2; Cohen's d = 0.51, p = 0.0005). This reduction was evident in both LTLE (d = 0.78, p = 0.0006) and RTLE (d = 0.31, p = 0.04).

Additionally, the contralateral entorhinal cortex (Cohen's d = 0.37, p = 0.01) also had reduced cortical thickness, albeit to a lesser degree, in those individuals with SE. There was no evidence of thinning in any other cortical region relating to a history of SE. Full regional results are presented in Supplementary Analysis 1.

In those individuals without a history of SE, the cortical thickness of both the ipsilateral (p = 0.54) and contralateral (p = 0.99) entorhinal cortices were not different to controls.

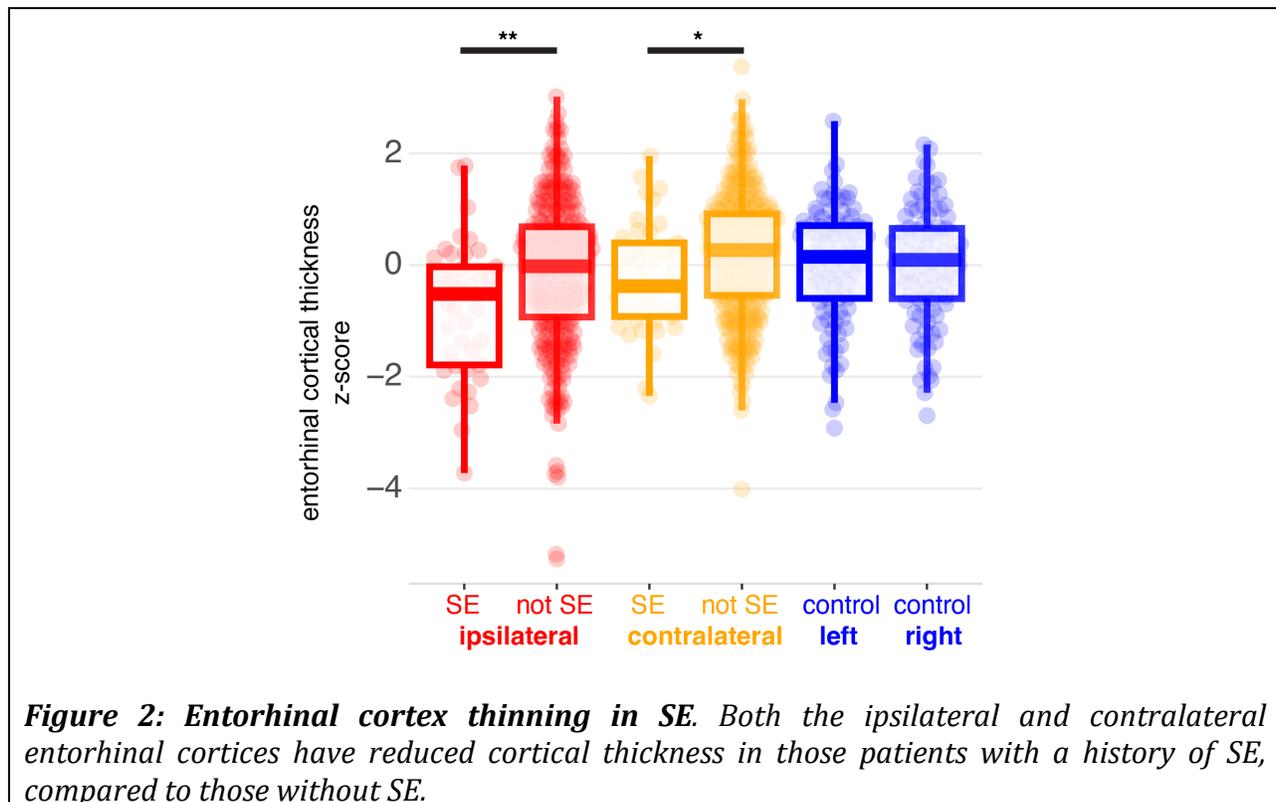

*Figure 2: Entorhinal cortex thinning in SE. Both the ipsilateral and contralateral entorhinal cortices have reduced cortical thickness in those patients with a history of SE, compared to those without SE.*

### Other structures are altered in TLE, but are unrelated to status epilepticus

Similar to previous results, we also found widespread cortical thinning in patients relative to controls, but this did not relate to SE. This thinning was most pronounced in bilateral precentral, paracentral, superior and inferior parietal regions (Cohen's d > 0.45, p < 0.001). Full regional results are presented in Supplementary Analysis 2.

### Hippocampal volume atrophy may be related to status epilepticus

In addition to cortical thickness, we also investigated the relationship between subcortical atrophy and history of SE. Compared to those without a history of SE, individuals with SE had reduced subcortical volume in the ipsilateral hippocampus (Cohen's d = 0.40, p = 0.02). There was no evidence of a similar effect in the contralateral hippocampus (p = 0.62), or any other subcortical region. Full regional results are presented in Supplementary Analysis 3.

Of the 41 individuals with TLE and SE in this cohort, the majority (31, 76%) had hippocampal sclerosis (HS), confirmed from postoperative histology. The remainder had dysembryoplastic neuroepithelial tumours (n=4), multiple pathologies (n=2), or some other non-specific pathology (n=4). Of the 316 individuals with TLE and no SE in this cohort, the majority (185, 59%) had HS. The remainder had dysembryoplastic neuroepithelial tumours (n=36), cavernoma (n=26), multiple pathologies (n=20). The remainder had other pathologies. The proportion of subjects with HS was significantly greater in those individuals with a history of SE ($\chi^2 = 3.74, p = 0.03$).

## Discussion

We investigated structural abnormalities in individuals with TLE and a history of SE after the onset of epilepsy, compared to those with epilepsy but no history of SE. We found atrophy of the entorinhal cortex both ipsilaterally and contralaterally. The effect was present irrespective of lateralisation and was most pronounced ipsilaterally, where it was equivalent to approximately 7% reduction in mean cortical thickness.

Given our cohort of exclusively TLE, other temporal structures may be relevant[12]. In particular,[5] highlight the role of other structures in seizure generation and continuation. Key structures highlighted include the hippocampus and amygdala, amongst others. We noted more extreme reductions in the subcortical volume of the ipsilateral hippocampus in those with a history of SE. Other brain regions, particularly in the parietal lobe, showed reduced cortical thickness in people with epilepsy compared to controls, but these reductions did not relate to a history of SE.

Does SE lead to neuronal damage? Longitudinal data with repeat scans post-SE may be revealing in this regard. In children,[3] found reduced hippocampal volumes in the 1-year followup data if hippocampal hyperintensities were observed in the acute phase post-SE. Additionally, individuals with SE (before habitual epilepsy onset) had reduced bilateral hippocampal volumes compared to those with only simple febrile seizures. In a case report,[13] described neocortical abnormalities which persisted in serial scans for at least 6-weeks post-SE.[14] analysed longitudinal data from 3 weeks to 12 months post-SE in nine adults. There, the authors reported volumetric alterations in patients relative to controls, but found no evidence for progressive changes. Taken together, these studies suggest that long-term atrophy may exist in individuals with a history of SE.

We performed our analysis on a large cohort available as part of the IDEAS data release[8]. This study serves as a demonstrative example of the utility of large datasets. Future studies should capitalise on such data to improve our understanding of other aspects of epilepsy, additional to sharing new data as recently suggested[15]. With prevalence of SE in our cohort of 11%, large efforts for multi-site data sharing would be advantageous to expand sample sizes and improve generalisability.

Our work has strengths and limitations. A strength is the availability of control data derived from the same scanners as the patients. This allows normative models to remove effects such as age and sex which are both known to influence cortical thickness[16].

Limitations of our work include potential bias in patient selection (all were undergoing presurgical evaluation, and proceeded to resections), and the age range limited to adults. Furthermore, we only explored cortical volumetrics and did not consider signal changes (e.g. as measured from T2 FLAIR MRI), previously reported to be affected by SE. Whilst we observed a difference in entorhinal cortex atrophy between those with and without a history of SE, the effect may have been greater in those with more frequent, or longer, instances of SE. Unfortunately, data for the frequency and duration of SE were not available.

In summary, we have shown an association between reduced cortical thickness in the ipsilateral and contralateral entorhinal cortex and a history of SE in a cohort of individuals with TLE.

## Acknowledgements

We thank members of the Computational Neurology, Neuroscience & Psychiatry Lab (www.cnnp-lab.com) for discussions on the analysis and manuscript; P.N.T. and Y.W. are both supported by UKRI Future Leaders Fellowships (MR/T04294X/1, MR/V026569/1). J.J.H. is supported by the Centre for Doctoral Training in Cloud Computing for Big Data (EP/L015358/1). JD, JdT are supported by the NIHR UCLH/UCL Biomedical Research Centre.